\documentstyle[12pt]{article}
\begin{document}

\baselineskip=0.7cm
\newcommand{\EQ}{\begin{equation}}
\newcommand{\EN}{\end{equation}}
\newcommand{\EQA}{\begin{eqnarray}}
\newcommand{\EQN}{\end{eqnarray}}
\newcommand{\EQAN}{\begin{eqnarray*}}
\newcommand{\EQNN}{\end{eqnarray*}}
\newcommand{\e}{{\rm e}}
\newcommand{\Sp}{{\rm Sp}}
\newcommand{\hzero}{{\hat{0}}}
\newcommand{\heleven}{{\hat{11}}}
\newcommand{\hr}{\sqrt{h}}
\newcommand{\tilded}{{\tilde{D}}}
\newcommand{\hPsi}{{\hat{\Psi}}}
\newcommand{\bu}{{(1)}}
\newcommand{\bl}{{(2)}}
\def\identity{{\rlap{1} \hskip 1.6pt \hbox{1}}}
\renewcommand{\theequation}{\arabic{section}.\arabic{equation}}
\newcommand{\Tr}{{\rm Tr}}
\renewcommand{\thesection}{\arabic{section}.}
\renewcommand{\thesubsection}{\arabic{section}.\arabic{subsection}}
\makeatletter
\def\section{\@startsection{section}{1}{\z@}{-3.5ex plus -1ex minus 
 -.2ex}{2.3ex plus .2ex}{\large}} 
\def\subsection{\@startsection{subsection}{2}{\z@}{-3.25ex plus -1ex minus 
 -.2ex}{1.5ex plus .2ex}{\normalsize\it}}
\def\appendix{
\par
\setcounter{section}{0}
\setcounter{subsection}{0}
\def\thesection{\Alph{section}.}}
\makeatother
\makeatletter
\@addtoreset{equation}{section}
\makeatletter
\def\thefootnote{\fnsymbol{footnote}}
\begin{flushright}
hep-th/0011122\\
UT-KOMABA/00-14\\
November 2000
\end{flushright}
\vspace{1cm}
\begin{center}
\Large
Supercurrents in Matrix theory and \\the generalized 
AdS/CFT correspondence

\vspace{1cm}
\normalsize
{\sc Yasuhiro Sekino}
\footnote{
e-mail address:\ \ sekino@hep1.c.u-tokyo.ac.jp}
\\
\vspace{0.3cm}
 {\it Institute of Physics\\
University of Tokyo, Komaba, Tokyo 153 }

\vspace{3.3cm}

Abstract\\

\end{center}
\noindent
We investigate Matrix theory in the large-$N$ limit 
following the conjectured correspondence between 
Matrix theory and  supergravity
on the near-horizon limit of the D0-brane background. 
We analyze the complete fermionic spectrum of supergravity
and obtain  two-point functions of the supercurrents
in Matrix theory. 
By examining the large-$N$ scaling properties of 
the correlators, we analyze the behavior of the supercurrents
under the boost in the 11-th direction and 
discuss the consistency of the 11-dimensional interpretation 
of the supersymmetry of Matrix theory.

\newpage
\section{Introduction}
According to the Matrix-theory conjecture of Banks, Fischler,
Shenker and Susskind \cite{bfss}, 
M-theory in the infinite momentum frame
is defined by the large-$N$ limit of 
the one-dimensional U($N$) super Yang-Mills theory
which describe the collection of $N$ D0-branes.
Since the D0-brane is believed to be
the Kaluza-Klein particle corresponding to the compactification 
of 11-th direction and has positive longitudinal momentum, 
it is considered as a natural candidate for the
elementary degrees of freedom in the infinite momentum
frame. Also, extended objects
of M-theory such as membranes and 5-branes can be
described as classical configurations of matrices
in SYM theory.
However, what is  highly non-trivial 
is that the SYM theory 
which is the description of D-branes in 
the low-energy, short-distance and the weak-coupling
 limit  
is proposed as 
the exact non-perturbative definition of M-theory.
Taking the large-$N$ limit must be
 an important ingredient.
Subsequently, it was proposed that
the finite-$N$ version of Matrix theory can be interpreted
as the compactification of M-theory on a lightlike circle 
\cite{susskind}. The  compactification in the lightlike direction
is 
equivalent to the compactification in the spacelike direction
with small compactification radius and small 
transverse length scale \cite{seiberg, sen}.  
Since M-theory compactified on the small spatial
circle is the weakly-coupled type IIA string theory,
it provides the explanation for the
appearance of SYM theory for the
D0-branes.
Various  quantitative checks for the finite-$N$ interpretation
has been performed, including the consistency with
the supergravity interactions up to two-loop
perturbations in Matrix theory \cite{bb, bbpt, oy}.

As Matrix theory is intrinsically formulated in 10-dimensional
language,
the crucial issue is
how it realizes the 11-dimensional
Lorentz invariance. In Matrix theory, 
the longitudinal momentum is given
by $P_{-}=N/R$ where $R=g_s\ell_s$ and the  
boost in the 11-th direction rescales $N$ with fixed
$g_s$, according to the proposal of Banks {\it et al}. 
Thus, for the boost invariance, the theory must have
a kind of scale invariance with respect to $N$ in the
large $N$ limit. However, analysis of gauge theory in the 
large $N$ limit (with fixed $g_s$) is difficult and there
was no clue for this `scale invariance'. 

In the previous paper \cite{sy}, with the motivation for
understanding the boost invariance of Matrix theory,
we have investigated Matrix theory in the 
large-$N$ limit using the conjectured 
correspondence with supergravity.
For non-dilatonic branes such as D3-branes, exact 
equivalence between super Yang-Mills theory describing
the branes and the superstring theory
on the near-horizon limit of the classical solution
is proposed \cite{maldacena}.
The same kind of correspondence is conjectured 
for dilatonic branes such as D0-branes \cite{imsy}. 
The conjecture was made
more precise by  
Jevicki and Yoneya \cite{jy} who pointed out that Matrix theory
and the near-horizon limit of the D0-brane solution
have the same symmetry
called the generalized conformal symmetry. 
In \cite{sy}, following this `generalized AdS/CFT correspondence'
and assuming the relation between the supergravity action 
evaluated on the near-horizon limit of D0-brane solution and 
the generating functional for Matrix-theory correlators, 
following the general method proposed by
 Gubser, Klebanov and Polyakov \cite{gkp} and
by Witten \cite{w} for non-dilatonic branes, 
we have predicted the 2-point
functions of bosonic operators of Matrix theory.
We have examined the $N$ dependence of the correlators
and analyzed the behavior of Matrix-theory operators
under the boost in the 11-th direction.
We have found that the scaling of $N$ 
can be interpreted as a spacetime symmetry, but there was an
unexpected
anomalous behavior for interpreting it as the Lorentz
boost. However,
this anomalous behavior does not necessarily mean
the violation of 11-dimensional Lorentz invariance 
in Matrix theory. 
We have to point out the possibility that the generalized
AdS/CFT correspondence does not give the exact information
for Matrix theory, which may be due to the fact that we have
to restrict ourselves to the near-horizon region 
of D0-branes in
the supergravity calculation in this formulation. 
We shall review these
discussions 
in section 4 of this paper for completeness.

In this paper, we study the Matrix-theory supercurrents.
The boost properties of the supercurrents are
particularly important, for they form the basis for
the following
11-dimensional interpretation of the supersymmetry of 
Matrix theory.
The SUSY in 11 dimensions has 32 real supercharges.
In the lightcone frame where SO(9) of SO(10,1)$\supset$
SO(1,1)$\times$SO(9) is manifest, supercharges are
divided into two classes 
which are called the kinematical SUSY and the dynamical SUSY,
according to the SO(1,1) weight
(longitudinal boost weight). 
The explanation for the 11-dimensional SUSY in Matrix theory
is as follows. Matrix theory has two kinds of fermionic 
symmetries: one is a constant shift of fermionic field only,
and the other is the usual SUSY transformation of SYM theory.
The former is interpreted as the kinematical SUSY and the
latter is interpreted as the dynamical SUSY in  11 dimensions.
Indeed, the Dirac-bracket algebra of the supercharges calculated
in Matrix theory reproduces the 11-dimensional
lightcone SUSY algebra under this identification
\cite{bss}.
In this paper, we try to
 give further evidence 
for the 11-dimensional SUSY of Matrix theory by checking
whether the supercharges (supercurrents) have the correct
SO(1,1) weight in the sense of the large-$N$ scaling
behavior. 
We continue the analysis of \cite{sy}
and compute the two-point functions of the supercurrents
of Matrix theory following the generalized AdS/CFT 
correspondence.
For that purpose, we work out the complete 
spectrum of the fermionic degrees of freedom of 10-dimensional
type IIA supergravity
on the near-horizon limit of D0-brane background.
Analysis of the spectrum is performed on the 
equivalent 11-dimensional background.
We analyze the $N$-dependence of the correlators 
which are obtained form the generalized AdS/CFT 
correspondence and 
discuss the behavior of the supercharges under the 
boost in the 11-th direction.

This paper is organized as follows. In section 2, 
we review the 10- and 11- dimensional descriptions
 of D0-brane solution.
In section 3, 
the fermionic spectrum of supergravity
is analyzed. In section 4,
we calculate Matrix-theory
two-point functions following the generalized AdS/CFT
correspondence. 
In section 5, we discuss the 11-dimensional Lorentz invariance
and supersymmetry of Matrix theory by analyzing the scaling
behavior of the correlators with respect to $N$.
We conclude in section 6.

\section{Ten- and eleven- dimensional descriptions of the solution}
\subsection{The D0-brane solution}
D0-brane solution in the string frame in 10 dimensions is given by
\EQA
&&ds_{10}^2=-\e^{2\tilde{\phi}/3}dt^2+\e^{2\tilde{\phi}/3}dx_m^2,\nonumber\\
&&\e^{\phi}=g_s\e^{\tilde{\phi}},\qquad A_0=-{1\over g_s}\left( {1\over 1+h}-1\right)\nonumber\\
&&\e^{\tilde{\phi}}=\left( 1+h\right)^{3/4},
\label{D0solution}
\EQN 
where
\[
h={q\over r^7},\quad q=60\pi^3(\alpha ')^{7/2}g_s N.
\] 
and $m(=1,\ldots, 9)$ denotes the indices of the 
Cartesian coordinates in the transverse directions
throughout this paper.
String coupling $g_s$ is defined by $\e^{\phi}$ at infinity
and the part of the dilaton which vanish at infinity
is denoted by $\tilde{\phi}$.

The D0-brane solution can be considered as the dimensional 
reduction of the solution in 11 dimensions. 10-dimensional
 description and 11-dimensional
description are equivalent if we note a few points remarked 
in the next subsection. We use 
the 11-dimensional description for the analysis of the 
spectrum in the next section, and we convert to the
10-dimensional language when studying the
generalized AdS/CFT correspondence which is formulated in 10
dimensions.
The metric in 11 dimensions
and the metric, dilaton and one-form field in 10 dimensions
are related by
\EQ
ds_{11}^2=\e^{-2\phi/3}ds_{10}^2+\e^{4\phi/3}
(dx^{11}-A_\mu dx^\mu)^2.
\label{dimred}
\EN
The metric which yields the D0-brane 
solution upon dimensional reduction on $x^{11}$ is given 
by\footnote{To derive (\ref{planewave}) from (\ref{D0solution}) using (\ref{dimred}),
we have rescaled the value of the coordinates
\[
x^{(11)}{}^\mu=g_s^{-1/3}x^{(10)}{}^\mu, \quad (\mu=0,\ldots 9 ),\quad
x^{(11)}{}^{11}=g_s^{2/3}x^{(10)}{}^{11}.
\label{rescalecoord}
\]
where $x^{(11)}{}^\mu$, $x^{(11)}{}^{11}$ 
are the coordinates used in (\ref{planewave}).
With the rescaling of $x^\mu$, 
 asymptotic flat 11D metric corresponds to 
asymptotic flat 10D metric. 
The rescaling of the 11-th coordinate is such that the period for 
the compactification is $R=g_s^{3/2} \ell_P=g_s\ell_s$ 
(not $=\ell_P$)
for $x^{(11)}$.}
\EQ
ds^2=dx^+ dx^- +h\; dx^- dx^-+dx_m^2
\label{planewave}
\EN
where $x^\pm=\pm t +x^{11}$.
%
We choose the following form of vielbein. 
\EQ
e^{\hzero}{}_{+}={1\over 2 \hr},\quad e^{\heleven}{}_{+}={1\over 2 \hr},\quad
e^{\hzero}{}_{-}=0,\quad e^{\heleven}{}_{-}=\hr
\label{vielbein}
\EN
Indices with hat refers to those for the local Lorentz frame
throughout this paper.
We summarize the connections and the curvatures of this
background in Appendix A.
\subsection{Near-horizon limit}
The 11-dimensional geometry given by (\ref{planewave}) 
corresponds to 
the full D0-brane solution (without taking near-horizon limit).
The near-horizon limit ($r\ll q^{1/7}$) of the D0-brane solution 
in 10 dimensions
 is given by substituting $h$ for $1+h$ in (\ref{D0solution}).
That metric is formally obtained from the metric 
\EQ
ds^2=2dt dx^- +h\; dx^- dx^-+dx_m^2
\label{nh11d}
\EN
by dimensional reduction on the $x^{11}$ direction.
(\ref{nh11d}) means that we can
 analyze the near-horizon limit of 
the 10-dimensional solution in the 11-dimensional language,
by reinterpreting the coordinate as follows.
\EQ
2t^{(n.h.)}=x^{(full)}{}^+,\quad x^{(n.h.)}{}^-=x^{(full)}{}^-
\label{reinterpretation}
\EN
where $x^{(full)}$ is the coordinate which was used in the last
subsection. 
We use the coordinate $x^{(full)}$ in the calculation
and make this reinterpretation at the end.

There is one thing which needs care. We want to analyze the spectrum of 
the fluctuations which {\it does not depend on the 11-th coordinate},
for it is directly related to the 10-dimensional interpretation.
Here, 11-th coordinate mean $x^{(n.h.)}{}^{11}$ when we want
to study the 
near-horizon limit. 
As the fluctuations are independent of $x^{(n.h.)}{}^{11}=
{1\over 2} x^{(full)}{}^+ +
x^{(full)}{}^-$ (but depends on $t^{(n.h.)}={1\over 2}x^{(full)}{}^+$ 
and other transverse coordinates), in the intermediate stage where we 
are calculating using the `({\it full})'coordinate,  we must take the fluctuations to be {\it independent of $x^{(full)}{}^-$}.\footnote{Compactification
of the supergravity solution (\ref{planewave}) 
on the lightlike circle is also discussed in \cite{hyun1,hyun2,hyun3}.}

\section{Fermionic spectrum on the near-horizon D0-brane background}

\subsection{Equations of motion and SUSY invariance}
We shall analyze the spectrum of the fermionic fluctuations
on the background (\ref{planewave}).
Fermionic fields of the 11-dimensional supergravity are the
 Rarita-Schwinger fields $\Psi_M$ which are Majorana
 spinors having vector indices. (We use $M=0,1,\ldots,9,11$ for
the indices of the 11-dimensional coordinate throughout this paper.)
Linearized equation of motion for the RS field is
\EQ
\Gamma^{MNP}D_N \Psi_P=0
\label{eom}
\EN
where $D_M$ is the covariant derivative containing
the torsionless spin connection at
 the linearized level.
Equations of motion for each component is given as follows.
We use the spherical coordinates
on the transverse $S^8$, which is denoted by $x^i$ ($i=1,\ldots,
8$). In the following equations, we use the 
`covariant derivative on the unit sphere' $\tilde{D}_i$ 
which is made from the connections on $S^8$. Covariant derivative
 $D_i$ on the background (\ref{planewave}) is not equal to $\tilde{D}_i$,
for $D_i$ has extra contributions from 
$\Gamma^r{}_{ij}$,
$\Gamma^i_{rj}$ and $\omega_{i\hat{r}\hat{\jmath}}$.
Note that the background (\ref{planewave}) is not a direct product of a sphere
and something. (See Appendix A.)
The $(M=+)$ component of (\ref{eom}) is 
\EQA
\Gamma^{+-}\Gamma^r(D_-\Psi_r -D_r\Psi_-)
-\Gamma^{+-}\Gamma^i\tilded_i\Psi_-  -{4\over r}\Gamma_r\Gamma^{+-}\Psi_- 
-\Gamma^+\Gamma^r\Gamma^i\tilded_i\Psi_r  -{4\over r}\Gamma^+\Psi_r \nonumber\\
-\Gamma^+\tilded^i\Psi_i
+\Gamma^{+-}D_-\Gamma^i\Psi_i +\Gamma^+\Gamma^r D_r\Gamma^i\Psi_i
+\Gamma^+\Gamma^j\tilded_j\Gamma^i\Psi_i +{9\over 2r}\Gamma^+\Gamma_r\Gamma^i\Psi_i=0
\label{zeroplus1}
\EQN
The ($M=-$) component is
\EQA
\Gamma^{-+}\Gamma^r(D_+\Psi_r -D_r\Psi_+)
-\Gamma^{-+}\Gamma^i\tilded_i\Psi_+  -{4\over r}\Gamma_r\Gamma^{-+}\Psi_+ 
-\Gamma^-\Gamma^r\Gamma^i\tilded_i\Psi_r  -{4\over r}\Gamma^-\Psi_r \nonumber\\
-\Gamma^-\tilded^i\Psi_i
+\Gamma^{-+}D_+\Gamma^i\Psi_i +\Gamma^-\Gamma^r D_r\Gamma^i\Psi_i
+\Gamma^-\Gamma^j\tilded_j\Gamma^i\Psi_i +{9\over 2r}\Gamma^-\Gamma_r\Gamma^i\Psi_i=0
\label{zerominus1}
\EQN
The ($M=r$) component is 
\EQA
\Gamma^{+-}(D_+\Psi_- -D_-\Psi_+)-\Gamma^+\Gamma^i\tilded_i\Psi_+
+{4\over r}\Gamma_r\Gamma^+\Psi_+ -\Gamma^-\Gamma^i\tilded_i\Psi_- 
+{4\over r}\Gamma_r\Gamma^-\Psi_-\nonumber\\
-\tilded^i\Psi_i
+\Gamma^+D_+\Gamma^i\Psi_i+\Gamma^-D_-\Gamma^i\Psi_i
+\Gamma^j\tilded_j\Gamma^i\Psi_i +{7\over 2r}\Gamma_r\Gamma^i\Psi_i=0
\label{zeror1}
\EQN
The ($M=i$) component contracted with $\Gamma^i$ (after
subtraction of a 
certain combination of (\ref{zeroplus1}), 
(\ref{zerominus1}) and (\ref{zeror1})) is
\EQA
-\Gamma^+\Gamma^i\tilded_i\Psi_+ +{4\over r}\Gamma_r\Gamma^+\Psi_+
-\Gamma^-\Gamma^i\tilded_i\Psi_- +{4\over r}\Gamma_r\Gamma^-\Psi_-
-\Gamma^r\Gamma^i\tilded_i\Psi_r -{4\over r}\Psi_r -2\tilded^i\Psi_i
\nonumber\\
+\Gamma^+D_+\Gamma^i\Psi_i +\Gamma^-D_-\Gamma^i\Psi_i 
+\Gamma^rD_r\Gamma^i\Psi_i
+ 2\Gamma^j\tilded_j \Gamma^i\Psi_i +{8\over r}\Gamma_r\Gamma^i\Psi_i=0
\label{zeroi1}
\EQN
The `$\Gamma^i$-transverse' part of ($M=i$) component
(which vanishes when contracted with $\Gamma^i$) is
\EQ
-\Gamma^\alpha \tilded_{(i)}\Psi_\alpha +\Gamma^\alpha D_\alpha\Psi_{(i)}
-\Gamma^r \tilded_{(i)}\Psi_r +\Gamma^r D_r\Psi_{(i)}
+\Gamma^j \tilded_{j}\Psi_{(i)}+{4\over r}\Gamma_r \Psi_{(i)}=0
\label{gammaitrsv}
\EN
where the subscript $(i)$ means that appropriate term proportional to
$\Gamma^i$ is substracted to make the expression $\Gamma^i$-transverse. 

The equations of motion are invariant under the
local supersymmetry under which the linearized fluctuations 
transform as
$
\delta\Psi_M=D_M \epsilon.
$
For general modes,
we fix this gauge invariance  
by imposing the condition
\EQ
\Gamma^i \Psi_i=0.
\label{generalgauge}
\EN
For the modes with lowest angular momentum,
the above condition cannot be imposed since
some of the components of the fluctuations are 
invariant under the SUSY transformation, as we
shall see in section 3.4.
That case will be treated separately.
 
We take
 the following explicit representation for the 
$\Gamma$ matrices in 11 dimensions which
have 32 $\times$ 32 components.
\EQ
\Gamma^\hzero =i \sigma^1\otimes \gamma^9,\;
\Gamma^{\hat{r}} =\sigma^2\otimes \gamma^9\;,
\Gamma^\heleven =\sigma^3\otimes \gamma^9\;,
\Gamma^{\hat{\imath}} =1\otimes \gamma^{\hat{\imath}}
\label{gamma11d}
\EN
where 
$\gamma^{\hat{\imath}}$ are 
Gamma matrices in 8 dimensions (16 $\times$ 16 matrices) 
which satisfy $\{ \gamma^{\hat{\imath}},
\gamma^{\hat{\jmath}}\}=2\delta^{\hat{\imath}\hat{\jmath}}$
and $\gamma^9=\gamma^1 \cdots\gamma^8$.

\subsection{Expansions into spinor and vector-spinor spherical harmonics}
Since the background has SO(9) symmetry
in the transverse space, it is appropriate to expand each field
into spherical harmonics which transform irreducibly under
SO(9).

We decompose the spinor fields on $S^8$ ($\Psi_{+},\Psi_{-}$ and $\Psi_{r}$) 
into spinor spherical harmonics ($\Xi^{k,+}$, $\Xi^{k,-}$) on $S^8$
which are 16-component Majorana spinors.
The coefficients of the expansion ($\hPsi_{+},\hPsi_{-}$ and $\hPsi_{r}$)
are 2-component spinors.
\EQA
&&\Psi_{+}=\sum_{+,-}\sum_{k=0}^\infty {\hat{\Psi}}^{k,\pm}_{+} (\gamma^9)^{-1/2} \Xi^{k,\pm},\quad
\Psi_{-}=\sum_{+,-}\sum_{k=0}^\infty {\hat{\Psi}}^{k,\pm}_{-} (\gamma^9)^{-1/2} \Xi^{k,\pm},\nonumber\\
&&\Psi_{r}=\sum_{+,-}\sum_{k=0}^\infty {\hat{\Psi}}^{k,\pm}_{r} (\gamma^9)^{-1/2} \Xi^{k,\pm}
\label{harms}
\EQN
Spinor spherical harmonics are 
the eigenfunctions of the Dirac operator
on the sphere
\EQ
\gamma^i\tilde{D}_i \Xi^{k,\pm}=\mp i (k+4)\Xi^{k,\pm}.\qquad (k=0,1,\ldots)
\label{eigenvaluexi}
\EN
As we can see from (\ref{eigenvaluexi}), $\Xi^{k,+}$ and $\Xi^{k,-}$
transform to each other when multiplied by $\gamma^9$.
 We have to take a linear combination
of $\Xi^{k,+}$ and $\Xi^{k,-}$ in the expansion (\ref{harms}) for
diagonalizing the equations of motion. 
(We have denoted the combination
by $(\gamma^9)^{-1/2}$ following \cite{nieuw}. $(\gamma^9)^{-1/2}$
has the property
$
\gamma^i (\gamma^9)^{-1/2}=-i (\gamma^9)^{1/2}\gamma^i.
$) 
The explicit forms of the harmonics $\Xi^{k,\pm}$ are given by
\EQ
\Xi^{k,\pm}=\left[ (k+7\pm i\gamma^i\tilde{D}_i)Y^k\right] \eta^{\pm}
\label{explicitxi}
\EN
where $Y^k$ is the scalar spherical harmonics of order $k$ 
which satisfy $\tilded^2 Y^k=-k(k+7)Y^k$. 
$\tilde{D}_i$ acts only on $Y^k$ in (\ref{explicitxi}). 
$\eta^{\pm}$ are the Killing spinors 
on $S^8$ which satisfy
\EQ
\tilded_i \eta^{\pm}=\mp {i\over 2}\gamma_i\eta^\pm
\label{killingspinor}
\EN
The $k=0$ modes of the spinor spherical harmonics
are the Killing spinors themselves.

We expand the vector-spinor on $S^8$, $\Psi_i$ which satisfy $\Gamma^i\Psi_i=0$
as 
\EQ
\Psi_i=\sum_{+,-}\sum_{k=1}^\infty \hPsi^{k,\pm}(\gamma^9)^{-1/2}\Xi^{k,\pm}_i
+\sum_{+,-}\sum_{k=1}^\infty \hat{\hPsi}{}^{k,\pm} (\gamma^9)^{-1/2} 
\tilded_{(i)}\Xi^{k,\pm}
\label{harmvs}
\EN 
The vector spinor harmonics $\Xi^{k,+}_i$, $\Xi^{k,-}_i$ 
are 16-component
Majorana spinors and have the properties
\EQ
\gamma^i \Xi_i^{k,\pm}=0,\quad \tilded^i\Xi_i^{k,\pm}=0
\EN
and are the eigenfunctions of the Dirac operator on the sphere
\EQ
\gamma^j\tilded_j\Xi^{k,\pm}_i=\mp i (k+4) \Xi^{k,\pm}_i\qquad (k=1,2,\ldots)
\EN
$\tilded_{(i)}\Xi^{k,\pm}$ is defined by 
\EQ
\tilded_{(i)}\Xi^{k,\pm}=\tilded_{i}\Xi^{k,\pm}\pm {i\over 8}(k+4)\gamma_i\Xi^{k,\pm}\qquad (k=1,2,\ldots)
\label{dxi}
\EN
and is the `$\gamma^i$-transverse' part of $\tilded_i\Xi$. 
(satisfies $\gamma^i\tilded_{(i)}\Xi=0$).
The $k=0$ mode is an exception.   
Since the vector-spinor harmonics are defined for $k\ge 1$
and $\tilded_{(i)}\Xi^{k,\pm}$
vanishes for $k=0$ (as we can see from (\ref{killingspinor}) and
(\ref{dxi})), $\Psi_i$ which satisfy $\gamma^i\Psi_i=0$
does not have the $k=0$ mode. However, we cannot impose the 
condition $\gamma^i\Psi_i=0$ in this case (as we will see in section 3.4), and we expand 
$\Psi_i$ for the $k=0$ mode without that condition as
\EQ
\Psi_i=-\sum \hPsi^{0,\pm}\gamma^i (\gamma^9)^{1/2}\eta^\pm.
\label{harmzerovs}
\EN

Since the spherical harmonics of different kind, or with
different
$k$, $+,-$ are orthogonal to each other, we can separately
analyze each mode.
The superscript $(k,+),(k,-)$ on the spherical harmonics will be
often omitted when there is no source of confusion.


\subsection{Transverse vector-spinor modes}
\label{spectrumtvs}
First, we analyze the mode corresponding to the vector-spinor 
harmonics $\Xi_i^{(k,\pm)}$.
We impose the gauge condition $\Gamma^i\Psi_i=0$
and substitute the harmonic expansions
(\ref{harms}) and (\ref{harmvs}) into the equations of 
motion. The vector-spinor modes only appear in
(\ref{gammaitrsv}) where the coefficient of $(\gamma^9)^{-1/2}
\Xi^{k,\pm}_i$ is
\EQ
\left[ 2\hr i \sigma^1\partial_+ +{1\over \hr}(-i\sigma^1+\sigma^2)\partial_-+\sigma^3\partial_r +{h'\over 4h} +{4\over r}\sigma^3 \mp{1\over r}(k+4)\right]\hPsi=0.
\label{eqi1}
\EN
where $h'=\partial_r h=-7 h/r$.

To analyze the near-horizon limit of the D0-brane solution,
we reinterpret $x^+\rightarrow 2t$ and assume the
fields to be independent of $x^-$, as explained in 
section 2.2.
Then, we change the radial variable to 
\EQ
z={2\over 5}q^{1/2}r^{-5/2}
\EN
and Euclidize the time coordinate 
($t=-i\tau$, $\omega_{M}=i\omega_{E}$).
\EQ
\left[ -\sigma^1\partial_\tau-\sigma^3\partial_z+(\mp{2\over 5}(k+4)+{7\over 10}){1\over z}\right](z^{-8/5}\hPsi)=0
\label{eucliddirac}
\EN

The solution which is regular at $r\rightarrow 0$ ($z\rightarrow\infty$) is given by
\EQ
\hPsi=\int {d\omega\over 2\pi}\e^{-i\omega\tau}z^{21\over 10}
\left\{ C^{(1)}(\omega)K_{{2\over 5}|{1\over 2}\mp(k+4)|}(|\omega|z)
+ C^{(2)}(\omega)K_{{2\over 5}|3\mp(k+4)|}(|\omega|z)\right\}
\label{isolution1}
\EN
where the superscript $(1)$ denotes the upper component of the
two component spinors and $(2)$ denotes the lower component,
($\sigma^3 C^{(1)}=+ C^{(1)}$ and $\sigma^3 C^{(2)}=-C^{(2)}$).
To be a solution of (\ref{eucliddirac}), $C^{(1)}$ 
and $C^{(2)}$ must satisfy
\EQ
i\omega \sigma^1 C^{(1)} =|\omega |C^{(2)}.
\label{isolution2}
\EN
Note that as is the case for the 1st-order action,
half of the components are physical and 
we can only fix either $\hPsi^{(1)}$ or $\hPsi^{(2)}$
by a boundary condition. 

\subsection{Spinor modes at the lowest level $(k=0)$}
Next, we analyze the spinor modes on the sphere 
with lowest angular momentum, $\Xi^{k=0,\pm}$.
First of all, we shall discuss the supersymmetry on the
D0-brane background.
The transformation $\delta \Psi_M =D_M\epsilon$
is written in components as 
\EQA
&&\delta \Psi_+=\partial_+\epsilon,\quad
\delta \Psi_-=\partial_-\epsilon+{h'\over 4\hr}\left(
(i\sigma^1-\sigma^2)
\otimes \identity\right) \epsilon,\nonumber\\
&&\delta \Psi_r=\partial_r \epsilon +{h'\over 4h} \left(
\sigma^3 \otimes \identity\right) \epsilon,\quad
\delta \Psi_i=\tilded_i \epsilon -{1\over 2}\left( \sigma^3
\otimes \gamma^9\gamma_i \right)\epsilon
\label{gaugecomp}
\EQN
When $\epsilon$ is in the $k=0$ mode (proportional to the 
Killing spinor)
\EQ
\epsilon=\hat{\epsilon}^\pm(\gamma^9)^{-1/2}\eta^\pm,
\EN
we can see from the property of the Killing spinor
(\ref{killingspinor})  that
$\Psi_i$ transforms as
\EQ
\delta\Psi_i=-{1\over 2}(\pm 1-\sigma^3)\hat{\epsilon}^\pm
\otimes\gamma_i (\gamma^9)^{1/2}\eta^\pm.
\label{zerogauge}
\EN
That is, half of the components of $\Psi_i$
are invariant under the 
SUSY and we cannot achieve $\Gamma^i\Psi_i=0$ for 
the $k=0$ modes.
We shall first analyze the equations for the 
$k=0$ modes without gauge 
fixing and
later specify a gauge conditions which are convenient.

We have to analyze four equations (\ref{zeroplus1}) $\sim$
(\ref{zeroi1}). (`$\Gamma^i$-transverse part (\ref{gammaitrsv})
is identically zero for $k=0$ modes.)
Substituting the harmonic expansions (\ref{harms}) and
(\ref{harmzerovs}) for $\Psi_+$,
$\Psi_-$, $\Psi_r$ and $\Psi_i$,
we obtain the following equations for $k=0$ modes
without gauge fixing.
The upper signs are for $\Xi^{k=0,+}$ mode and the 
lower signs are for $\Xi^{k=0,-}$.
From (\ref{zeroplus1}),
\EQA
{i\over \hr} \sigma^1(\partial_-\hPsi_r-\partial_r\hPsi_-)
-{h'\over 4h^{3/2}}\sigma^2 \hPsi_-+{4\over \hr r}(-i\sigma^1\pm\sigma^2)\hPsi_-+{4\over r}(-1\pm \sigma^3)\hPsi_r\nonumber\\ +{h'\over 4h}(-1+\sigma^3)\hPsi_r
-{8\over \hr r}\sigma^2\partial_-\hPsi -{8\over r}\sigma^3\partial_r\hPsi
-{2 h'\over hr}\sigma^3\hPsi-{28\over r^2}(\pm 1+\sigma^3)\hPsi=0 
\label{zeroplus}
\EQN
From (\ref{zerominus1}),
\EQA
&&2(\partial_+\hPsi_r-\partial_r\hPsi_+)-{h'\over 2h}\sigma^3
\hPsi_++
{8\over r}(\pm\sigma^3 -1)\hPsi_++(1\pm 1){4\over \hr r}
(i\sigma^1 -\sigma^2)\hPsi_r -{16\over r}\sigma^3\partial_+\hPsi\nonumber\\
&&-{8\over \hr r}(i\sigma^1 -\sigma^2)\partial_r\hPsi+{2 h'\over h^{3/2} r}
(i\sigma^1-\sigma^2)\hPsi+(-1\pm 1){28\over \hr r^2}(i\sigma^1-\sigma^2)\hPsi=0
\label{zerominus}
\EQN
From (\ref{zeror1}),
\EQA
&&\hspace{-1.2cm}2\sigma^3(\partial_-\hPsi_+-\partial_+\hPsi_-)+{h'\over 2h}(i\sigma^1-\sigma^2)\hPsi_++{8\hr\over r}(\pm i\sigma^1-\sigma^2)\hPsi_+
 +(1\pm 1){4\over \hr r}(-i\sigma^1+\sigma^2)\hPsi_-\nonumber\\
&&-{16\hr\over r}i\sigma^1 \partial_+\hPsi -{8\over \hr r}
(-i\sigma^1+\sigma^2)
\partial_-\hPsi -{28\over r^2}(\pm 1+\sigma^3)\hPsi=0 
\label{zeror}
\EQN
From (\ref{zeroi1}),
\EQA
-{\hr\over r}(\mp i\sigma^1+\sigma^2)\hPsi_+ +(1\pm 1){1\over 2\hr r}(-i\sigma^1+\sigma^2)\hPsi_- -{1\over 2 r}(1\mp\sigma^3)\hPsi_r\nonumber\\
 -{2\hr\over r}i\sigma^1\partial_+ \hPsi -{1\over \hr r}(-i\sigma^1+\sigma^2)\partial_-\hPsi-{1\over r}\sigma^3\partial_r\hPsi-{h'\over 4 h r}\hPsi-{7\over r^2}(\pm1+\sigma^3)\hPsi=0 
\EQN
We have 8 first-order 
equations (four 2-component equations) for
$2\times 4$ variables ($\hPsi_+$, $\hPsi_-$, $\hPsi_r$, 
$\hPsi$).
As is the case for Rarita-Schwinger field, $3\times 2$ 
degrees of freedom are reduced by gauge fixing
and constraints and remaining degrees of freedom are
halved due to the Dirac constraint. Thus, there is
one physical degree of freedom. We have obtained the
Klein-Gordon-type 2nd-order equation for physical
mode which should correspond to the result of elimination
of unphysical variable from Dirac-type 1st-order equations.

The physical mode corresponding to
 $\Xi^{k=0,+}$ (the upper signs in the above equations)
satisfies 
\EQ
\left[-\partial_t^2+\partial_z^2+{1\over z}\partial_z 
-({21\over 5})^2{1\over z^2}\right](z^{-7/2} 
\hPsi_-^{(1)})=0
\label{physical0plus}
\EN
where $\hPsi_-^{(1)}$ is the upper component of $\hPsi_-$
($\sigma^3 \hPsi_-^{(1)}=+\hPsi_-^{(1)}$). 
The derivation of (\ref{physical0plus}) is presented in
Appendix B. We remark here that we have done the
reinterpretation $x^+\rightarrow 2t$ and set $\partial_-=0$
which corresponds to considering the near-horizon limit
in 10 dimensions.
Of course, our choice of the physical degree of freedom 
($\hPsi_-^{(1)}$)
is not unique for it can be expressed by (the combination of) other
variables using the constraint equations. 
In addition,
 we have to note that the order of the Bessel function
$\nu$ ($=21/5$ in this case) is not unique. That is, when there is
a field which satisfy Bessel equation with order $\nu$
\EQ
\left[-\partial_t^2+\partial_z^2+{1\over z}\partial_z 
-{\nu^2\over z^2}\right]\phi=0,
\EN
The following combination satisfies
the Bessel equation with the order shifted by $\pm 1$.
\footnote{The difference of the orders for the 
upper ($C^{(1)}$) and the lower ($C^{(2)}$)
component of the vector-spinor mode $\hPsi$ in 
(\ref{isolution1}) is an example of this ambiguity
of the order of the physical degree of freedom.}
\EQ
\left[-\partial_t^2+\partial_z^2+{1\over z}\partial_z 
-{(\nu\pm 1)^2\over z^2}\right](\partial_z\phi\mp {\nu\over z}\phi)=0.\EN
This fact corresponds to the identity of the Bessel function
\EQ
zK'_\nu(z)\pm \nu K_\nu(z)=-zK_{\nu\mp 1}(z).
\EN
In (\ref{physical0plus}), 
we have chosen the representative
for physical mode whose order $\nu$ allows
interpretation in terms of 
 the generalized AdS/CFT correspondence,
as we shall see in section 4.

As a result of a similar analysis for $\Xi^{k=0,-}$ mode
(the lower signs in the field equations),
we obtain the physical mode for this case
\EQ
\left[-\partial_t^2+\partial_z^2+{1\over z}\partial_z 
-({14\over 5})^2{1\over z^2}\right](z^{-7/2}\Psi_-^{(2)})=0.
\label{physical0minus}
\EN
where $\sigma^3 \hPsi_-^{(2)}=-\hPsi_-^{(2)}$. 
The same remark as in the $\Xi^{k=0,+}$ case for
the order of the Bessel function applies.

\subsection{Spinor modes at general level $(k=1,2,\ldots)$ }
\label{spectrumgeneral} 

In this case, we analyze 5 equations (\ref{zeroplus1})
 $\sim$ (\ref{gammaitrsv}).
Under the gauge condition
$\Gamma^i\Psi_i=0$, the equations for $\Xi^{k,\pm}$ modes $(k=1,2,\ldots)$
are the following. The upper signs are for the
$\Xi^{k,+}$ mode and
the lower signs are for the $\Xi^{k,-}$ mode. 
From (\ref{zeroplus1}),
\EQA
&&-2(\partial_-\hPsi_r-\partial_r\hPsi_-)
+{h'\over 2h}\sigma^3\hPsi_- +{2\over r}\left(\mp(k+4)\sigma^3+4
\right)\hPsi_-
\nonumber\\
&&\hspace{-1cm}
+{h'\over 2\hr}(-i\sigma^1+\sigma^2)\hPsi_r -{2\hr\over r}\left(
\mp(k+4)\sigma^2+4i\sigma^1\right)\hPsi_r
+{7\over 4}k(k+8){\hr\over r^2}
i\sigma^1\hat{\hPsi}=0
\label{gplus}
\EQN
From (\ref{zerominus1}),
\EQA
&&2(\partial_+\hPsi_r-\partial_r\hPsi_+)
-{h'\over 2h}\sigma^3\hPsi_+ +{2\over r}\left(\pm(k+4)\sigma^3-4
\right)\hPsi_+
\nonumber\\
&&+{1\over r\hr}\left(
\mp(k+4)-4\right)(-i\sigma^1+\sigma^2)\hPsi_r
+{7\over 8}k(k+8){1\over r^2\hr}
(-i\sigma^1+\sigma^2)\hat{\hPsi}=0
\label{gminus}
\EQN
From (\ref{zeror1}),
\EQA
&&-2\sigma^3(\partial_+\hPsi_--\partial_-\hPsi_+)
-{h'\over 2\hr}(-i\sigma^1+\sigma^2)\hPsi_+ 
+{2\hr \over r}\left(\pm(k+4)i\sigma^1-4\sigma^2\right)\hPsi_+
\nonumber\\
&&+{1\over r\hr}\left(\pm(k+4)+4\right)(-i\sigma^1+\sigma^2)\hPsi_-
+{7\over 8}k(k+8){1\over r^2}\hat{\hPsi}=0
\label{gr}
\EQN
From (\ref{zeroi1}),
\EQA
&&2\hr\left(\mp(k+4)i\sigma^1+4\sigma^2\right)\hPsi_+
+{1\over \hr}\left(\mp(k+4)-4\right)(-i\sigma^1+\sigma^2)\hPsi_-
\nonumber\\
&&+\left(\mp(k+4)\sigma^3+4\right)\hPsi_r
-{7\over 4}k(k+8){1\over r}\hat{\hPsi}=0
\label{ga}
\EQN
From (\ref{gammaitrsv}),
\EQA
&&-2\hr i\sigma^1\hPsi_+ -{1\over \hr}(-i\sigma^1+\sigma^2)\hPsi_- 
-\sigma^3\hPsi_r +2\hr i\sigma^1\partial_+\hat{\hPsi}\nonumber\\
&&+{1\over \hr}(-i\sigma^1+\sigma^2)\partial_-\hat{\hPsi}
+\sigma^3\partial_r \hat{\hPsi} +{h'\over 4h}\hat{\hPsi}
+{3\over 4}\left(\mp(k+4)+4\sigma^3\right){1\over r}\hat{\hPsi}=0
\label{gti}
\EQN

There are two independent physical degrees of freedom.
This is as expected, for the total number of degrees of freedom
for $k\ge 1$ mode equals that of the RS field. That is,
$2\times 16$ for spinor modes ($\Xi$) plus $(8-2)\times 16$ for 
vector-spinor modes ($\Xi_i$) 
($-2$ for the conditions $D^i\Xi_i=\gamma^i\Xi_i=0$) 
equals the total degrees of freedom of RS fields $(11-3)\times 16$.
We shall explain the outline of the calculation of the 
spectrum and present the results.
First, we note that there are 5 sets of equations for 4 sets
of variables. We can see that the system of the equations
is consistent 
by noticing that (\ref{gti}) is obtained by taking a
linear combination of derivatives of 
other four equations. Thus, 
we consider only the first four of the equations.
Setting $\partial_-=0$ (and reinterpreting $x^+\rightarrow 2t$), 
we can eliminate $\hPsi_i$, $\hPsi_r$ and $\hPsi_+$ algebraically
using (\ref{ga}), (\ref{gr}) and (\ref{gplus})
and obtain the following equation for $\hPsi_-$.
\EQA
&&-{2\over 7}h\partial_t^2\hPsi_-+{2\over 7}\partial_r^2\hPsi_-
+\left\{ {32\over 7}+\left(\mp{4\over 7}(k+4)-1\right)\sigma^3
\right\}{1\over r}\partial_r\hPsi_-\nonumber\\
&&+\left\{ {2\over 7}(k+4)^2\pm(k+4)+{753\over 56}
+\left( \mp{30\over 7}(k+4)-{15\over 2}\right)\sigma^3\right\}
{1\over r^2}\hPsi_-+{\left(\mp(k+4)+4\sigma^3\right)\over k(k+8)}
\nonumber\\
&&\cdot
\left\{\left({8\over 7}(k+4)^2\pm 2(k+4)-{58\over 7}\right)
i\sigma^1+\left(\mp 2(k+4)-10\right) \right\}{\hr\over r}\partial_+
\hPsi_-=0
\label{secondcouple}
\EQN
After diagonalizing this second-order coupled equations for 
two variables ($\hPsi^{(1)}_-$ and $\hPsi^{(2)}_-$ where
$\sigma^3\hPsi^{(1)}_-=+\hPsi^{(1)}_-$ and 
$\sigma^3\hPsi^{(2)}_-=-\hPsi^{(2)}_-$), we  
found two physical 
modes which are solved by the modified Bessel functions
of the following order $\nu$.
For $\Xi^{k,+}$ modes,
\EQA
\nu_1={2\over 5}k+{21\over 5} \qquad(k=0,1,\ldots),\nonumber\\
\nu_2={2\over 5}k+{7\over 5} \qquad(k=1,2,\ldots).
\label{ordergplus}
\EQN
where we have included the case $k=0$
in the formula.
Explicit forms of the diagonalized fields $\varphi_1^{k,+}$ and
$\varphi_2^{k,+}$ corresponding to $\nu_1$ and $\nu_2$ respectively
are given (for $k\ge 1$) by 
\EQ
\varphi_1^{k,+}=
2i{(k+8)\over (2k+9)} 
z^{-5/2}\left\{\partial_z\hPsi_-^{(1)}
+\left({2\over 5}k-{21\over 10}\right){1\over z}
\hPsi_-^{(1)}\right\}
+z^{-5/2}\partial_t\hPsi_-^{(2)}
\EN
and
\EQ
\varphi_2^{k,+}=2i{(k+8)\over (2k+9)} 
z^{-5/2}\left\{\partial_z\hPsi_-^{(1)}
-{({31\over 10}k+{119\over 5})\over
(k+8)z}
\hPsi_-^{(1)}\right\}
+z^{-5/2}\partial_t\hPsi_-^{(2)}.
\EN

For $\Xi^{k,-}$ modes, the orders of the Bessel 
functions are 
\EQA
\nu_1={2\over 5}k \qquad(k=1,2,\ldots),\nonumber\\
\nu_2={2\over 5}k+{14\over 5} \qquad(k=0,1,\ldots).
\label{orderminus}
\EQN
Explicit forms of the diagonalized fields ($\varphi_1^{k,-}$
and $\varphi_2^{k,-}$)
for $k\ge 1$ are
obtained from the following expressions. 
\EQA
\tilde{\varphi}_1^{k,-}&=&
{2ik\over (2k+7)} 
z^{-5/2}\left\{\partial_z\hPsi_-^{(1)}
-\left({2\over 5}k+{53\over 10}\right){1\over z}
\hPsi_-^{(1)}\right\}
+z^{-5/2}\partial_t\hPsi_-^{(2)},\nonumber\\
\tilde{\varphi}_2^{k,-}&=&{2ik\over (2k+7)} 
z^{-5/2}\left\{\partial_z\hPsi_-^{(1)}
-{({31\over 10}k+1)\over kz}
\hPsi_-^{(1)}\right\}
+z^{-5/2}\partial_t\hPsi_-^{(2)}.
\label{soltilde}
\EQN
$\tilde{\varphi}_1^{k,-}$ and
$\tilde{\varphi}_2^{k,-}$  in (\ref{soltilde}) 
 are the solutions with orders
$\tilde{\nu}_1=2k/5 -1$ and $\tilde{\nu}_2=2k/5+9/5$ respectively
which are related to the orders $\nu$
given in (\ref{orderminus}) by $\nu=\tilde{\nu}+1$.
Thus, we can construct the solutions with order $\nu$ using
the method described in the last 
subsection.\footnote{Different expressions for $\varphi_1^{k,-}$, 
$\varphi_2^{k,-}$ having the orders (\ref{orderminus})
 may be possible by using (\ref{secondcouple}) again.}
The orders $\nu$ of (\ref{orderminus}) 
are the ones which allow the interpretations
in terms of the generalized AdS/CFT correspondence.


\section{Generalized AdS/CFT Correspondence}
\subsection{Matrix-theory correlators  form supergravity}
As we did in the case of bosons, we shall calculate the correlators 
in Matrix theory by evaluating the supergravity action.
First, we recall in what situations the calculation of
classical supergravity is valid \cite{sy}. 
The curvature of the D0-brane background must be small
in string unit which gives
\EQ
r\ll (g_s N)^{1/3}\ell_s
\EN
and the effective string coupling (dilaton background)
must be small
\EQ
g_s^{1/3} N^{1/7}\ell_s\ll r.
\EN
Those two conditions  are simultaneously satisfied in the whole
near-horizon region
\EQ
r\ll (g_s N)^{1/7}\ell_s, 
\EN
if  
\EQ
N\rightarrow\infty,\quad (g_s N)>1.
\label{sugravalid}
\EN

 
We assume the following relation between 
the supergravity action and the generating functional of the Matrix-theory
correlators.
\EQ
\e^{-S_{SG}[\Psi]} = \langle \exp 
\{\int dt \sum_I \Psi_I (t){\cal O}_I(t)\}
\rangle ,
\label{cprescription}
\EN
where we impose the boundary conditions on the physical modes of 
supergravity at the end of the near-horizon region ($z=q^{1/7}$
{\it i.e.}  
$r\propto (g_s N)^{1/7}\ell_s$)
and evaluate the action on the near-horizon limit
of the D0-brane background
as a functional of the boundary conditions. 
In this paper, we obtain the two-point functions of the fermionic operators
of Matrix theory.


We first analyze the transverse vector-spinor mode 
with the `$+$' choice of the eigenvalue of the 
Dirac operator on the sphere ($\Xi_i{}^{(k,+)}$ mode)
and then present the results for general modes.
The relevant part of the 10-dimensional action becomes (up to
a numerical
factor,)
\EQ
S=-i {1\over \kappa^2}\int dt dr r^8\sum_I {\hat{\Psi}}^I{}^\dagger (i\sigma^1)
\left\{ i\sigma^1 \hr \partial_t +\sigma^3\partial_r +\cdots \right\}
{\hat{\Psi}}^I.
\EN
We have used the orthonormality of the spherical harmonics
\EQ
\int d\Omega_8 g^{ij} {\Xi^I_i}^\dagger \Xi^J_j =C \delta^{IJ}
\EN
where $C$ is a numerical constant and $I,J$ are the labels
for the harmonics. 
Also note that we have chosen the normalization ($g_s$ dependence)
of the
fields such that the
action in 10 dimensions
has $\kappa^2
\sim 1/g_s^2\ell_s^8$ as a prefactor.

%
%
Euclidizing the time coordinate ($t=-i\tau$) and  using the radial
coordinate $z=2q^{1/2}r^{-5/2}/5$,   
\EQ
S= {1\over \kappa^2}\int d\tau dz q^{8/5}z^{-16/5}
\sum_I {\hat{\Psi}^I}{}^\dagger (\sigma^1)
\left\{ \sigma^1 \partial_\tau +\sigma^3\partial_z +\cdots \right\}
{\hat{\Psi}^I}
\EN
As noted in section 3.3, we can impose the boundary condition
to one of the two components of $\hPsi$. Following the choice
which is assumed in the case of ordinary AdS/CFT correspondence
\cite{hs}, we fix the component which is more divergent as
we take the boundary to infinity ($z\rightarrow 0$).
That is, we fix $\hPsi^{(1)}$ for the $\Xi_i^{k,+}$ mode
(by noting $K_\nu(z)\sim z^{-\nu}$). For the same reason,
we fix $\hPsi^{(2)}$ in the case of the $\Xi_i^{k,-}$ mode. 
The action is of first order 
and the bulk contribution vanishes on shell, but 
we need to add the following boundary term  
(at $z=q^{1/7}$) to the action 
\EQ
S=S_{boundary}=-{1\over 2\kappa^2}q^{8/5-16/35}\int d\tau \hat{\Psi}^\dagger
\sigma^1\hat{\Psi}
\label{boundaryaction}
\EN
to ensure that the classical solution is 
really an extremum of the action 
 when we fix $\hPsi^{(1)}$ at the boundary
\cite{henneaux}.
%

Substituting the 
 solution for the $\Xi_i^{(k,+)}$ mode 
\EQ
\hPsi=\int {d\omega \over 2\pi} \e^{-i\omega\tau} \left( z/q^{1/7}\right)^{21/10}
\left\{ K_{{2\over 5}(k+{7\over 2})}
(|\omega | z)+ {i\omega\over |\omega|}\sigma^1
K_{{2\over 5}(k+1)}(|\omega| z)\right\}
{\hPsi^{(1)}_b(\omega)\over K_{{2\over 5}(k+{7\over 2})}
(|\omega| q^{1/7})}
\EN
which satisfy
boundary condition
$\hPsi^{(1)}(\tau,z=q^{1/7})=\int d\omega 
\e^{-i\omega\tau}\hPsi^{(1)}_b(\omega)/2\pi$,
the action reads
\EQA 
S&=&{1\over \kappa^2}q^{8/7}\int {d\omega \over 2\pi} \hat{\Psi}^{(1)}_b
(-\omega)
\hat{\Psi}^{(1)}_b(\omega) {i\omega \over |\omega |} {K_{\nu -1}(|\omega | q^{1/7}) \over K_{\nu}(|\omega| q^{1/7})}\nonumber\\
&=& 2^{-2\nu+1}{\Gamma(1-\nu) \over \Gamma(\nu)} {1\over \kappa^2}q^{1+2\nu/7}\int {d\omega \over 2\pi}\hat{\Psi}^{(1)}_b (-\omega)
\hat{\Psi}^{(1)}_b (\omega)i\omega\left\{ |\omega |^{2\nu-2} +\cdots\right\}
\label{actionvalue1}
\EQN
where 
\[
\nu={2\over 5}k+{7\over 5}
\]
and we have retained only the leading part which 
is non-analytic in $\omega$. We have used the Majorana
condition (for our representation, $\hPsi^\dagger=\hPsi^T 
\sigma^3$).

Following the ansatz of the generalized AdS/CFT correspondence 
(\ref{cprescription}), we obtain 
the two-point function of the operator ${\cal O}$ which couple to the
mode $\Psi^{(1)}$ 
\EQA
\langle {\cal O}(\omega){\cal O}(-\omega)\rangle&=&-{\delta\over 
\delta\hat{\Psi}_b^{(1)}(-\omega)}
{\delta\over \delta\hat{\Psi}_b^{(1)}(\omega)} S_{SG}\nonumber\\
&=& -{2^{-2\nu+1}\over \pi}{ \Gamma(1-\nu)\over \Gamma(\nu)}
{1\over \kappa^2}q^{1+2\nu/7} i\omega  
\left\{|\omega|^{2\nu-2}+\cdots\right\}.
\EQN
Fourier transforming to the position space, 
\EQA
\langle {\cal O}(\tau_1){\cal O}(\tau_2)\rangle
&=&\int {d\omega\over 2\pi}\e^{-i\omega\tau}\int {d\omega'\over 2\pi}\e^{-i\omega'\tau}\langle {\cal O}(\omega){\cal O}(\omega')\rangle\nonumber\\
&=&{1\over 2\pi^{-3/2}}{\Gamma(\nu+1/2)\over \Gamma(\nu)}
 {1\over \kappa^2}q^{1+2\nu/ 7}
{(\tau_1-\tau_2)\over |\tau_1-\tau_2|^{2\nu+1}}
\label{correlator}
\EQN
The 2-point functions corresponding to the other physical
modes are given by substituting into 
(\ref{correlator}) the order $\nu$ 
obtained in sections 3.3, 3.4 and 3.5.

\subsection{Identification of the Matrix-theory operators}
Following the generalized AdS/CFT 
correspondence (\ref{cprescription}),
we have calculated
 the 2-point functions of Matrix-theory operators
which couple to Rarita-Schwinger fields of the 
11-dimensional supergravity.
Now we would like to identify the 
operators which couple to each 
physical mode of supergravity. 
RS fields are the gauge fields of 
the supersymmetry and should couple to the supercurrents 
of Matrix theory, but we do not have principles
which determine the coupling. We shall determine the
Matrix-theory operator by requiring 1) the 
scaling dimension (with respect to the
generalized conformal symmetry which is explained below)
agrees with
the one read off from the correlator
which was obtained form the supergravity calculation,
and 2)
 the representation with respect to the transverse rotation group
SO(9) agrees with the supergravity mode.
Note that this situation is essentially the same as in the 
ordinary AdS/CFT correspondence such as the $AdS_5/CFT_4$ case,
where the representation
with respect to the ${\cal N}=4$ superconformal symmetry
is the guideline for determining the coupling.

Matrix theory is defined by the following action.
\EQ
S={1\over g_s \ell_s}\int dt \Tr \left({1\over 2}D_t X^m
D_t X^m +{i\over 2}\theta D_t \theta +{1\over 4\ell_s^4}[X^m,X^n]^2
-{1\over 2\ell_s^2}\theta \gamma^m [\theta,X^m]\right)
\label{matrixaction}
\EN
where $D_t=\partial_t +i[A,\quad]$.
and $\gamma^m$ are the 9-dimensional Gamma matrices which 
are real and symmetric. 
It has the generalized conformal symmetry which allows
the coupling constant $g_s$ to be transformed. The scale
transformation is given by
\EQ
t\rightarrow \lambda^{-1} t,\quad X_m\rightarrow \lambda X_m,
\quad \theta\rightarrow\lambda^{3/2}\theta,\quad
g_s\rightarrow \lambda^{3}g_s.
\label{gcscale}
\EN
Together with the time translation and the special conformal
transformation (treating as $g_s$ as a field with dimension:3),
they constitute the conformal algebra. The near-horizon limit
of the D0-brane solution has the same kind of
symmetry \cite{jy}.

Matrix theory has two kinds of fermionic symmetries.
One is a constant shift of fermionic field only.
\EQ
\delta \theta= \tilde{\epsilon}
\label{matrixkinematicalsusy}
\EN 
The other is the usual supersymmetry of the SYM theory.
\EQA
&&\delta X^m= i \epsilon^T \gamma^m \theta,\quad
\delta A= i \epsilon^T \theta,\nonumber\\
&&\delta \theta= -{1\over 2}\left( 
2D_t X_m\gamma^m +{i\over \ell_s^2}[X_m,X_n]\gamma^{mn}\right)
\epsilon
\label{matrixdynamicalsusy}
\EQN

The explicit forms of the 
supercurrents of Matrix theory are obtained by 
Taylor and van Raamsdonk \cite{tvr}.
They are identified by reinterpreting the one-loop 
effective potential between a pair of diagonal blocks
in Matrix theory as the potential between the currents
in the linearized supergravity compactified on $x^-$.
The ($M=+$) and ($M=m$) components of the supercurrents are 
given as follows up to numerical constant factors and 
possible three-fermion terms. The currents are assumed
to be integrated in the $x^-$-direction.
\EQA
\tilde{q}^+&=& {1\over R}\Tr \left( \theta \right)\nonumber\\
\tilde{q}^m&=& {1\over R}\Tr 
\left( \{\dot{X}{}^m-F_{mn}\gamma^{n}\}
\theta \right)\nonumber\\
q^+&=&{1\over R}\Tr \left( \{\dot{X}{}^m\gamma_m
+{1\over 2}F_{mn}\gamma^{mn}\}\theta\right)\nonumber\\
q^m&=&{1\over R}{\rm STr} \left( \{2\dot{X}{}^m\dot{X}{}^n
\gamma_n+\dot{X}{}^m F_{np}\gamma^n\gamma^p
+2\dot{X}{}_n F_{pm} \gamma^n\gamma^p
+F_{np}F_{qm}\gamma^n\gamma^p\gamma^q\}\theta\right)
\label{matrixcurrents1}
\EQN
where $F_{mn}=i[X_m,X_n]$ and
${\rm STr}$ is the symmetrized trace in which the
 average is taken over  all possible
orderings of matrices (treating $F_{mn}$
as a single unit). The above expression is for the $A=0$
gauge and we have set $\ell_s=1$. Note that $\tilde{q}^+$
and $q^+$ agree with the supercharges in the SYM theory
corresponding to (\ref{matrixkinematicalsusy}) 
and (\ref{matrixdynamicalsusy}).
The ($M=-$) components of the supercurrents
are of the form (up to 3-fermion terms)
\EQA
\tilde{q}^-&=& {1\over R}{\rm STr} \left( \{ \mbox{terms of 2nd 
order 
in }F_{mn}, \dot{X}_m\}\theta \right)\nonumber\\
q^-&=&{1\over R}{\rm STr} \left( \{\mbox{terms of 3rd order 
in }F_{mn}, \dot{X}_m\}\theta \right)
\label{matrixcurrents2}
\EQN
To obtain the explicit forms of $\tilde{q}^-$ and
$q^-$, 
analysis of the
`$1/r^8$ terms' of Matrix-theory effective 
potential is needed. (Terms in (\ref{matrixcurrents1})
are obtained from the `$1/r^7$ terms'.)
In addition to the operators (\ref{matrixcurrents1})
and (\ref{matrixcurrents2}), there are the `moments' of the
currents which couple to the transverse derivatives of the
supergravity fields. The $k$-th moment $q_{(k)}^{M\{m_1,\ldots,m_k\}}$,
$\tilde{q}_{(k)}^{M\{m_1,\ldots,m_k\}}$ contain 
a term which is obtained by inserting $k$ times 
the transverse field $\tilde{X}{}^m=X^m/q^{1/7}$ into the traces
in (\ref{matrixcurrents1}) and (\ref{matrixcurrents2}).

The consistency of these couplings between the branes
and the supergravity fields are confirmed in various contexts:
they are consistent with the couplings obtained by 
expanding the Matrix-regularized membrane action in the lightcone
gauge around the flat background \cite{dnp};
the couplings for D0-branes which are obtained through the
Seiberg-Sen argument are consistent with the Born-Infeld action \cite{tvr2};
the couplings for D$p$-branes obtained from D0 couplings by T-duality
indeed have $(p+1)$-dimensional Lorentz covariance \cite{tvr3}; 
and notably, the leading part of the absorption
cross sections by D3-branes in the small $(g_s N)$, large $N$ limit
which is calculated in the $D=4$, ${\cal N}=4$ SYM theory using the 
T-dualized couplings
exactly reproduce the ones calculated in the semiclassical supergravity
including the cases of higher partial waves \cite{ktr}.
 
We assume that these operators are the candidates
for the operators which couple to the supergravity
modes in our case. 
We shall compare the 
scaling dimensions (with respect to the generalized
conformal symmetry (\ref{gcscale})) of the supercurrents and that 
of the correlators obtained in the last subsection.
From the right hand side of (\ref{correlator}),
we see that 
the scaling dimension $\Delta$ of the operator is
\EQA
\Delta&=&{1\over 2}\left\{3\cdot(-2)+3\cdot (1
+{2\nu\over 7})+ 2\nu\right\}\nonumber\\
&=&-{3\over 2}+{10\over 7}\nu
\label{dim}
\EQN

We can consistently identify the Matrix-theory operator 
which have the dimension (\ref{dim}) for every
supergravity modes. The results are
summarized in the Table below. 
Operators $q^m_{(k)}$ and $\tilde{q}^m_{(k)}$ in the Table
are the ones which have the same SO(9) quantum numbers
as the harmonics. {\it e.g.}, $\tilde{q}_{(1)}^{m\{ m_1\}}
\sim {1\over R}{\rm STr}\{(\dot{X}^m-F_{mn}\gamma^n)X^{m_1}\theta\}
-(m\leftrightarrow m_1)$.
\vspace{1cm}

\begin{center}
{\bf Table }\nopagebreak\\
Physical modes of supergravity and the corresponding 
Matrix-theory operators
\nopagebreak
\vspace{0.3cm} 
\[
\begin{array}{|c||c|c|c|c|c|c|}
\hline
\mbox{SUGRA fields}&\multicolumn{2}{|c|}{\Psi_i}&\multicolumn{4}{|c|}
{\Psi_+,\Psi_{-},\Psi_r, \Psi_i}\\
\hline
\mbox{harmonics}&\Xi_i^{k,+}&\Xi_i^{k,-}&\Xi^{k,+}&\Xi^{k,+}
&\Xi^{k,-}&\Xi^{k,-}\\
\hline
\mbox{physical modes}& z^{-7/2}\hPsi^\bu&z^{-7/2}\hPsi^\bl
&\varphi_1^{k,+}&\varphi_2^{k,+}&\varphi_2^{k,-}&\varphi_1^{k,-}\\
\hline
\mbox{order} \, \, \nu&{2\over 5}k+{7\over 5}&{2\over 5}k+{14\over 5}
&{2\over 5}k+{21\over 5}&{2\over 5}k+{7\over 5}&
{2\over 5}k+{14\over 5}&{2\over 5}k\\
\hline
\mbox{regions of }k &k\ge1 &k\ge1 &k\ge 0 &k\ge 1
&k\ge 0 &k\ge 1\\
\hline
\mbox{operator }{\cal O}& \tilde{q}^m_{(k)} & 
q^m_{(k)} & q^-_{(k)} & q^+_{(k)} & 
\tilde{q}^-_{(k)} & \tilde{q}^+_{(k)} \\
\hline 
\mbox{dimensions of }{\cal O}&{1\over 2}+{4\over 7}k&{5\over 2}+{4\over 7}k&
{9\over 2}+{4\over 7}k&{1\over 2}+{4\over 7}k&{5\over 2}+{4\over 7}k&
-{3\over 2}+{4\over 7}k\\
\hline
\end{array}
\]
\end{center}
\vspace*{0.5cm}

\section{11-dimensional Lorentz invariance and Supersymmetry of Matrix theory}
In this section we consider the implications of our results
for the Matrix-theory conjecture, 
following the line of arguments 
of the previous paper \cite{sy}. Some of the results of \cite{sy}
are reviewed for completeness.
According to the conjecture of Banks, Fischler,
Shenker and Susskind, infinite momentum frame ($P_-\rightarrow
\infty$) is achieved by taking the limit 
\EQ
N\rightarrow\infty\qquad \mbox{with}\quad g_s :\mbox{ fixed}.
\label{imlimit}
\EN
In this region, Matrix theory is expected to have the 
boost invariance in the $x^{11}$ direction,
which rescales $N$ with fixed $g_s$.
As argued in section 4.1, our results obtained from supergravity
is valid when $N\rightarrow\infty$ and
$(g_s N)>1$, which will allows us to investigate the infinite
momentum limit.\footnote{Strictly 
speaking, the condition for the validity of 
supergravity (\ref{sugravalid})
is different from the infinite momentum limit (\ref{imlimit}).
To satisfy (\ref{sugravalid}),
we need to take $g_s\rightarrow 0$ as $N\rightarrow\infty$ 
rather than fixing $g_s$. We are assuming that this difference does not 
affect the results.}
We shall check the boost invariance in the sense of 
the large-$N$ scaling behavior of the operators.
 
Under the boost transformation
\EQ
x^+\rightarrow \e^\omega x^+,\quad
x^-\rightarrow \e^{-\omega} x^-,
\label{boost1}
\EN
where $P_-$ scales as $N$,
the time coordinate of Matrix theory
which is the lightcone time scales as $N$, Matrix-theory
Hamiltonian ($P_+$) must scale as $1/N$ while  
the transverse coordinates and the compactification
radius are kept fixed.
We shall analyze the behavior of the operators under
the boost from the scaling behavior of the correlators.
We substitute $\tau\rightarrow N\tau$ in the correlators
and examine the $N$ dependence.
\EQ
\langle {\cal O}(\tau_1) {\cal O}(\tau_2)\rangle \rightarrow
N^{2 d_{IMF}}G(\tau_1 -\tau_2) 
\EN
where 
$G(\tau)$ is a function independent of $N$ and
we have defined the `scaling dimension with respect to 
the boost transformation'
$d_{IMF}$ of the operator ${\cal O}$. 

First, we review the behavior of the correlators of the bosonic
operators.  From the generalized AdS/CFT
correspondence we have obtained \cite{sy}
\EQ
\langle {\cal O}(\tau_1){\cal O}(\tau_2) \rangle 
= {1\over g_s^2\ell_s^8}(g_s N\ell_s^7)^{1+2\nu/ 7}
{1\over |\tau_1-\tau_2|^{2\nu+1}}.
\label{bose1}
\EN
where $\nu$ is the order of the modified Bessel function
of the corresponding supergravity mode. By examining
the values of $\nu$,
we have found that the `dimension' $d_{IMF}$ is 
given by
\EQ
d_{IMF}= {6\over 7}\nu=(1+{1\over 5})(n_+ -n_- -1)-({1\over 7}+{1\over 5})k.
\EN
where $n_+ (n_-)$ is the number of the upper $+(-)$ index on the
operator and $k$ means the $k$-th moment.
The fact that $d_{IMF}$ is 
determined
solely from the spacetime index structure of the operators
suggests that the scaling of $N$ is indeed related to the
spacetime symmetry. The tensor with the $+$ index indeed scales
inversely as the tensor with the $-$ index. However, 
the weights $\pm 6/5$ are
different from the natural expectation $\pm 1$
by $\pm 1/5$.
The interpretation of the remaining terms are as follows.
The $k$-th moment has a contribution
$-(1/7+1/5)k$. The factor $-{1/7}$ is explained by the fact
that the moments which we are considering are made by multiplying 
$\tilde{X}^m\sim X^m/((g_s N)^{1/7} \ell_s)$ 
 (but not $X^m$) $k$ times, but there is also an
 anomalous behavior $-1/5$ for each transverse field.
The constant part ($-(1+1/5)$) is interpreted
as coming from the $x^-$ integration when we define the currents,
also with the anomalous behavior for $x^-$ ($-1/5$).

Now we shall analyze the scaling behavior of the
supercurrents.
From the correlator obtained in the last section 
\EQ
\langle {\cal O}(\tau_1){\cal O}(\tau_2) \rangle 
= {1\over g_s^2\ell_s^8}(g_s N\ell_s^7)^{1+2\nu/ 7}
{(\tau_1-\tau_2)\over |\tau_1-\tau_2|^{2\nu+1}},
\label{fermi1}
\EN
we find the following scaling properties
of the fermionic operators
of Matrix theory.
\EQA
d_{IMF}&=& 
{1\over 2}-{6\over 7}\nu\nonumber\\
&=& (1+{1\over 5})(n_+-n_--1)-({1\over 7}+{1\over 5})k\mp 
{1\over 2}(1+{1\over 5})-{1\over 10}
\label{fermidimf}
\EQN
where the ($-$) sign on ${1\over 2}(1+{1\over 5})$ is for
the current $q^M$ and the (+) sign is for the current
$\tilde{q}^M$. 
The interpretations of the scaling for the lightcone indices,
for $x^-$-integration 
and for the moments are the same as in the bosonic case.
In addition, we see that $q^M$ and $\tilde{q}^M$ transform
inversely. This is indeed needed for spacetime
interpretation. The current $\tilde{q}^M$ is
the supercurrent corresponding to the constant shift 
of fermionic field (\ref{matrixkinematicalsusy})
in Matrix theory, which is interpreted
as the kinematical SUSY in 11 dimensions.  The current
$q^M$ is the supercurrent corresponding to the 
supersymmetry of the SYM theory (\ref{matrixkinematicalsusy}),
which is interpreted as the
dynamical SUSY in 11 dimensions.
The kinematical supercharge ($\tilde{q}$) 
satisfy $\Gamma^-\tilde{q}=0$ and 
the dynamical supercharge ($q$) 
satisfy $\Gamma^+ q=0$ to form the lightcone
SUSY algebra ($\{ \tilde{q},\tilde{q} \}\sim P_-\Gamma^-$,
$\{ \tilde{q},q \}\sim P_m\Gamma^m$,
$\{ q,q \}\sim P_+\Gamma^+$). 
Under the boost (\ref{boost1}),
the kinematical and dynamical SUSY charges transform as
\EQ
\tilde{q}\rightarrow \e^{-\omega\Gamma^{0,11}/2}\tilde{q}
=\e^{\omega/2}\tilde{q},\quad
q\rightarrow \e^{-\omega\Gamma^{0,11}/2}q=\e^{-\omega/2}q,
\label{lorentzqQ}
\EN
which can be shown from the properties of $q,\tilde{q}$
mentioned in the above sentences.
Thus we expect that 
the current $\tilde{q}^M$ has a contribution
$-1/2$ for the boost weight, and the current $q^M$
has $+1/2$.  Our results shows anomalous behavior
$\pm 1/10$ also in this case. We do not have
an interpretation for the last term $-1/10$ of
(\ref{fermidimf}).

The interpretation of the anomalous behavior is not clear.
One possibility is that Matrix theory does not have 11-dimensional 
Lorentz invariance or supersymmetry,
 but the following interpretation 
seems more appropriate \cite{sy}.
 We have studied Matrix theory by analyzing the 
supergravity inside the near-horizon region. However, it is possible 
that we could not take into account the necessary degrees of freedom
of Matrix theory by restricting to the near-horizon
region. Indeed, `average transverse size' of Matrix theory
$L^2\equiv {\rm Tr} (X^m X^m) /N$ is estimated to be $L\sim (g_s N)^{1/3}\ell_s$
\cite{polchinski},
which is larger than the `boundary' of the near-horizon region $r_b\sim
(g_s N)^{1/7}\ell_s$ in the case of interest $(g_s N)>1$.
Further argument supporting this interpretation was given in \cite{yoneya99}. 
Suppose we shift the boundary $r_b\sim N^{1/7}$ to 
some larger value $\tilde{r}_b\sim N^\alpha$.
Assuming that the system is homogeneous (though it will not really
be the case), the field $X^m$ 
will be rescaled to
 ${\tilde{r}_b\over r_b}X^m$. 
 If we require the factor just cancels
the anomalous behavior for transverse fields $N^{-1/5}$,
the shifted boundary must be 
\EQ
\tilde{r}_b\propto N^{1/7+1/5} =N^{1/3+1/105}
\EN
which is very close to $L\sim N^{1/3}$. It seems to
suggest that we need to have radial size at least 
$L\sim (g_s N)^{1/3}\ell_s$ to analyze Matrix theory.

\section{Discussions}
In this paper, we have predicted the 2-point functions of the
supercurrents of Matrix theory following the conjecture of
generalized AdS/CFT correspondence. 
We have analyzed the scaling
behavior with respect to $N$ and considered whether 
the Matrix-theory 
supercurrents have correct weights when we interpret the scaling
of $N$ (with fixed $g_s$) as the boost in the 11-th direction 
following the original proposal 
of Banks, Fischler, Shenker and Susskind
\cite{bfss}.
We have reached to a result which is not very definitive:
The supercurrents corresponding to the two kinds of 
fermionic symmetries
of Matrix theory can be
indeed interpreted as having inverse weights under 
the Lorentz boost,
which suggests the correctness of the interpretation
of them as kinematical and dynamical SUSY currents
of the 11-dimensional SUSY.
However, the weights 
are slightly different from the 
canonical behavior, which seem to be of the same nature
as the anomalous behavior 
for the spacetime tensors which were found in the previous
paper. The origin of these behavior are not clear but
it is likely that the generalized AdS/CFT correspondence
is  not able to give the exact information
of Matrix theory, possibly due to the
`IR cutoff' of the supergravity calculation
which is the boundary of the near-horizon region.

To understand the effect of this cutoff more precisely, 
we will have to study the cutoff dependence from both 
supergravity and gauge theory. On the supergravity side,
one possibility is to try to extend further the
generalized AdS/CFT correspondence 
assuming that the bulk/boundary correspondence 
continues to hold for outside the near-horizon region.
The holographic renormalization group \cite{holographicrg}
is a suitable framework for analyzing the
effect of shifting the position of the boundary.
On the gauge-theory side, on the other hand,
analyzing the dependence on
the cutoff (for transverse fields
$X^m$) may be as difficult as solving
the theory directly. 
Up to now, quantitative study of the large-$N$ Matrix theory
in the strong coupling is only
attempted through
the Gaussian approximation \cite{gaussian1,gaussian2}.
We hope that the cutoff dependence 
can be incorporated in some approximation scheme.

Though we have mentioned the possibility for the limitation
of the generalized AdS/CFT correspondence, 
our results which exhibit the large-$N$ scaling behavior
should be the consequence of 
the collective dynamics of Matrix theory.
Though the near-horizon region $r<(g_s N)^{1/7}\ell_s$ is
smaller than the average transverse size 
$L\sim (g_s N)^{1/3}\ell_s$, it should be noted that
it is larger than the characteristic size of $N$-body
bound state obtained by a mean field analysis \cite{bfss}
$L_{m.f.}\sim (g_s N)^{1/9}\ell_s$. 
We would like to note a fact
which may be an indication \cite{sy} that the generalized
AdS/CFT correspondence takes into account certain amounts of 
 degrees of freedom of Matrix theory.
Assume that the coefficient of the two-point function
of the stress-energy tensor gives the entropy of the
theory. The coefficient ($c$) of the two-point function
of $T_{ij}$ in the lowest angular momentum $(k=0)$ mode
which is obtained from the generalized
AdS/CFT correspondence 
$
c \propto N^2 (g_s N)^{-3/5}
$
agrees with the $g_s$ and $N$ dependence of the entropy
which is calculated from the Bekenstein-Hawking 
formula
$
S\propto N^2 (g_s N \ell_s^3/T_H^3)^{-3/5}
$
where $T_H$ is the Hawking temperature.
We hope to clarify the meaning of this agreement
in future works.

Once the nature and the validity of the generalized
AdS/CFT correspondence for Matrix theory are understood,
we can use this formalism to investigate other 
dilatonic D-branes, for the general D$p$-branes have
the generalized conformal symmetry.
Particularly interesting case is that of  D1-branes.
The U($N$) super Yang-Mills theory in 2 dimensions
describing D1-branes is called Matrix string
theory and is conjectured to be
another non-perturbative definition of the type IIA
string theory \cite{dvv}.
In the large-$N$ limit with small $g_s$, this theory 
is believed to be described effectively
by a conformal field theory, and an effective interaction
vertex operator is proposed which is claimed to have
leading scaling dimension.
The generalized AdS/CFT correspondence for D1-branes
will be useful in checking  these conjectures
and in investigating the non-perturbative behavior of 
 string theory.

\vspace{0.5cm}
\noindent
Acknowledgement

I would like to thank T. Yoneya for valuable
discussions at every stage of this work and for careful
reading of the manuscript. I also thank
W. Taylor for reading the manuscript and for giving
important comments.

\vspace{1cm} 
\noindent
{\large Appendix}
\renewcommand{\theequation}{\Alph{section}.\arabic{equation}}
\appendix 
\setcounter{equation}{0}
\vspace*{0.3cm}
\section{The connections and the curvatures of the background}
We summarize here the connections and the curvatures
of the background in 11 dimensions which is given
by the metric
\[
ds^2=dx^+ dx^- +h\; dx^- dx^-+dx_m^2.
\]
where $h=q/r^7$.
Using the spherical coordinate on $S^8$, the metric reads
\EQ
g_{++}=0,\; g_{+-}={1\over 2}, \;g_{--}=h ,\;g_{rr}=1,\;g_{ij}=r^2 k_{ij} 
\EN
where $k_{ij}$ is the metric of the unit sphere $S^8$
and $x^i$  ($i,j=1,\ldots 8$) are the (angular) 
coordinate along $S^8$. 

Non-vanishing components of the Christoffel symbols are 
\EQ
\Gamma^r{}_{--}=-{1\over 2} h',\;\Gamma^+{}_{-r}= h',\;
\Gamma^i{}_{jr}={1\over r} \delta^i_j,\;\Gamma^r{}_{ij}=-{1\over r} g_{ij},\;
\Gamma^i{}_{jk}=\Gamma^i{}_{jk}{}^{(S^8)}
\EN
where $\Gamma^i{}_{jk}{}^{(S^8)}$ means the Christoffel symbol on the
unit sphere $S^8$ and $h'=\partial_r h$.
Non-vanishing components of the curvature tensors are
\EQ
R^i{}_{-j-}=-{h'\over 2r}\delta^i_k={7\over 2}{q\over r^9}\delta^i_k,\quad R^r{}_{-r-}=-{1\over 2}h''=-28{q\over r^9}.
\EN
With the choice of the vielbein
\EQ
e^{\hzero}{}_{+}={1\over 2 \hr},\quad e^{\heleven}{}_{+}={1\over 2 \hr},\quad
e^{\hzero}{}_{-}=0,\quad e^{\heleven}{}_{-}=\hr,
\EN
non-vanishing components of the spin connections are
\EQ
\omega_{-\hzero\hat{r}}=-{h'\over 2\hr},\;\omega_{-\heleven\hat{r}}={h'\over 2\hr},\;\omega_{r\hzero\heleven}=-{h'\over 2 h},\;
\omega_{i\hat{r}\hat{\jmath}}=-{1\over r}e_{\hat{\jmath}i}e_{\hat{r}r},\;
\omega_{i\hat{\jmath}\hat{k}}=\omega_{i\hat{\jmath}\hat{k}}{}^{(S^8)}
\label{spinconnection}
\EN
where $\omega_{i\hat{\jmath}\hat{k}}{}^{(S^8)}$ means the spin connection
on the unit sphere $S^8$.

We define the covariant derivative which is made from the 
connections on $S^8$, $\tilded_i$. The covariant derivative
$D_i$ has extra contributions 
from 
$\Gamma^r{}_{ij}$,
$\Gamma^i_{rj}$, $\omega_{i\hat{r}\hat{\jmath}}$.
For example, the covariant 
derivative of a spinor $\epsilon$ is
\EQA
D_i\epsilon&=&\partial_i\epsilon +{1\over 4}\omega_{i\hat{M}
\hat{N}}\Gamma^{\hat{M}\hat{N}}\epsilon =\tilded_i\epsilon
+{1\over 2}\omega_{i\hat{r}
\hat{\jmath}}\Gamma^{{\hat{r}}\hat{\jmath}}\epsilon\nonumber\\
&=&\tilded_i\epsilon
 -{1\over 2}\left( \sigma^3\otimes \gamma^9\gamma_i
\right)\epsilon
\EQN
using the representation of the $\Gamma$ matrix (\ref{gamma11d}).

\section{Spectrum of the $\Xi^{k=0,+}$ mode}
In this appendix, we describe the details of the analysis of the 
spectrum of the $\Xi^{k=0,+}$ mode and show that there is one 
physical degree of freedom which is solved by the Bessel equation
with order $\nu=21/5$.
We have $2\times 4$ components of equations 
(\ref{zeroplus1}) $\sim$ (\ref{zeroi1}) 
 for 8 variables ($\hPsi_+^{(1)}$, $\hPsi_+^{(2)}$,
$\hPsi_-^{(1)}$, $\hPsi_-^{(2)}$, $\hPsi_r^{(1)}$, $\hPsi_r^{(2)}$,
$\hPsi^{(1)}$ and $\hPsi^{(2)}$, where the superscripts (1) and (2) denote
upper and lower component of the two-component spinors, respectively). 

There is  a local supersymmetry $\delta\hPsi_M=D_M\epsilon$
under which the fields transform as 
\EQA
&&\delta\hPsi_+=\partial_+\hat{\epsilon},\quad
\delta\hPsi_-=\partial_-\hat{\epsilon}+{h'\over 4\hr}(-\sigma^2+i\sigma^1)
\hat{\epsilon}\nonumber\\
&&\delta\hPsi_r=\partial _r\hat{\epsilon}+{h'\over 4h}\sigma^3\hat{\epsilon},
\quad\delta\hPsi={1\over 2}(1-\sigma^3)\hat{\epsilon}
\EQN
The parameter $\hat{\epsilon}$ is defined by 
$\epsilon=\hat{\epsilon}(\gamma^9)^{-1/2}\eta$ where $\eta$ is the 
Killing spinor on $S^8$.
We use half of the gauge freedom and set $\hPsi^{(2)}$=0. Other half
will be fixed in the following. The equations of motion in components
with the condition $\hPsi^{(2)}$=0 and $\partial_-=0$ are
the following. From (\ref{zeroplus1}),
\EQA
&&-{i\over \hr}\partial_r \hPsi_-^{(2)}+i({h'\over 4 h^{3/2}} -{8\over \hr r})
\hPsi_-^{(2)} -{8\over r}\partial_r \hPsi^{(1)} -({56\over r^2}+{2 h'\over rh})
\hPsi^{(1)}=0
\label{a1}\\
&&-{i\over \hr}\partial_r \hPsi_-^{(1)}-{i h'\over 4 h^{3/2}}\hPsi_-^{(1)}
-({8\over r}+{h'\over 2 h})\hPsi_r^\bl=0
\label{a2}
\EQN
From (\ref{zerominus1}),
\EQA
&&2\partial_+\hPsi_r^\bu-2\partial_r\hPsi_+^\bu -{h'\over 2 h}\hPsi^\bu_+
+{16 i\over r\hr }\hPsi_r^\bl -{16\over r}\partial_+\hPsi^\bu=0
\label{a3}\\
&&2\partial_+\hPsi_r^\bl-2\partial_r\hPsi_+^\bl 
+(-{16\over r}+{h'\over 2 h})\hPsi^\bl_+=0
\label{a4}
\EQN
From (\ref{zeroi1}),
\EQA
&&-2\partial_+\hPsi_-^\bu+i\hr ({16\over r}+{h'\over h})\hPsi_+^\bl
-{16i\over \hr r}\hPsi_-^\bl -{56\over r^2}\hPsi^\bu=0
\label{a5}\\
&& 2\partial_+\hPsi_-^\bl-{16 i \hr \over r}\partial_+\hPsi^\bu=0
\label{a6}
\EQN
From (\ref{gammaitrsv}),
\EQA
&&-{2i\over r\hr}\hPsi_-^\bl+{2i\hr\over r}\hPsi_+^\bl
-{1\over r}\partial_r\hPsi^\bu -({14\over r^2}+{h'\over 4rh})\hPsi^\bu=0
\label{a7}\\ 
&&-{1\over r}\hPsi_r^\bl-{2i\hr\over r}\partial_+\hPsi^\bu=0
\label{a8}
\EQN

To solve this set of equations, first note that (\ref{a6}) can be written as
\[
\partial_+ (\hPsi_-^\bl -{8i\hr \over r}\hPsi^\bu)=0
\]
and (\ref{a1}) can be written as
\[
\partial_r \{ r^8h^{-1/4}(\hPsi_-^\bl -{8i\hr \over r}\hPsi^\bu)\}=0,
\]
thus, we can set 
\EQ
\hPsi_-^\bl={8i\hr \over r}\hPsi^\bu.
\label{b1}
\EN
From (\ref{a8}), we find
\EQ
\hPsi_r^\bl=-2i\hr \partial_+\hPsi^\bu.
\label{b2}
\EN
From (\ref{a7}) and (\ref{b1}), 
\EQ
\hPsi_+^\bl=-{r\over 16h}\{ \partial_r \hPsi_-^\bl -({1\over r}+{h'\over 4h})
\hPsi_-^\bl\}
\label{b3}
\EN

Using (\ref{b1}) $\sim$ (\ref{b3}) to eliminate $\hPsi^\bu$, $\hPsi^\bl_r$
and $\hPsi^\bl_+$ (and using $h'=-7 h/r$), (\ref{a2}) can be written as
\EQ
i\partial_r (h^{1/4}\hPsi_-^\bu)={9\over 8}h^{3/4}\partial_+\hPsi_-^\bl
\label{dirac1}
\EN
and (\ref{a5}) can be written as 
\EQ
\partial_+\hPsi_-^\bu =-{9i \over 32\hr}\partial_r \hPsi_-^\bl
-{603 i\over 128 r\hr}\hPsi_-^\bl
\EN
This set of `Dirac equation' is solved by $\hPsi_-^\bu$ which satisfy
\EQ
[-\partial_t^2+\partial_z +{1\over z}\partial_z -({21\over 5})^2{1\over z^2}]
z^{-7/2}\hPsi_-^\bu=0
\EN
where we have reinterpreted $x^+\rightarrow 2t$ and changed the 
radial variables $z=2 q^{1/2}r^{-5/2}/5$.

We have found one physical degree of freedom $\hPsi_-^\bu$ and
we have seen that the variables $\hPsi^\bu$, $\hPsi^\bl_r$,
$\hPsi^\bl_+$ and $\hPsi_-^\bl$ are determined using the
equations (\ref{b1}) $\sim$ (\ref{dirac1}).
Remaining 2 variables, $\hPsi_+^\bu$ and $\hPsi_r^\bu$ are
determined by (\ref{a3}) and using the remaining gauge freedom
(to set {\it e.g.} $\hPsi_+^\bu=0$).
We can check that the set of solutions satisfy the equation (\ref{a4})
which we have not used for the derivation,
which shows that the system of the equations is consistent.

\end{document}